\documentclass[superscriptaddress,amsmath,floatfix,aps,prc,twocolumn]{revtex4-2}
\DeclareUnicodeCharacter{03C9}{$\omega$}
\DeclareUnicodeCharacter{03C3}{$\sigma$}

\usepackage{amssymb}  % \gtrsim, \geqslant, etc: see amsguide.ps
\usepackage{graphicx}
\usepackage{exscale}
\usepackage{textcomp}
\usepackage{enumerate}
\usepackage{amsmath}
\usepackage{amssymb}
\usepackage[compat=1.1.0]{tikz-feynhand} 
\usepackage{multirow}
\usepackage{orcidlink}
\usepackage{color}

\usepackage[normalem]{ulem}  % \sout{old text} for strikeout

\newcommand{\ga}{\gamma}
\newcommand{\Ga}{\Gamma}
\newcommand{\de}{\delta}
\newcommand{\De}{\Delta}
\newcommand{\ep}{\varepsilon}

\newcommand{\ka}{\kappa}

\newcommand{\om}{\omega}

\newcommand{\bea}{\begin{eqnarray}}
\newcommand{\eea}{\end{eqnarray}}
\newcommand{\ba}[1]{\begin{array}{#1}}
\newcommand{\ea}{\end{array}}

\newcommand{\beq}{\begin{equation}}
\newcommand{\eeq}{\end{equation}}

\newcommand{\eqn}[1]{(\ref{#1})}
\newcommand{\sliver}{\kern 0.07em} % like italic correction \/

\newcommand{\cm}{\text{cm}}

\renewcommand{\sec}{\text{s}}
\newcommand{\gram}{\text{g}}
\newcommand{\MeV}{\text{MeV}}
\newcommand{\kHz}{\text{kHz}}

\newcommand{\nsat}{\ensuremath{n_{0}}}
\newcommand{\tdamp}{\ensuremath{\tau_\text{damp}}}

\newcommand{\muIeq}{\ensuremath{\mu_I^\text{eq}}}

\newcommand{\chixI}{\raisebox{0.4ex}{$\chi$}_{x_I}}
\newcommand{\chinB}{\raisebox{0.4ex}{$\chi$}_{n_B}}

\newcommand{\nbar}{{{\bar n}_B}}
\newcommand{\half} {{\textstyle \frac{1}{2}}}
\newcommand\deriv[3]{{\dfrac{\partial #1}{\partial #2}}\biggr|_{#3}}
\newcommand{\p}{\partial}

\begin{document}

\title{Isospin Equilibration in Neutron Star Mergers}
\author{Mark G.~Alford\,\orcidlink{0000-0001-9675-7005}}
\email{alford@wustl.edu}
\affiliation{Physics Department, Washington University in Saint Louis, 63130 Saint Louis, MO, USA}

\author{Alexander Haber\,\orcidlink{0000-0002-5511-9565}}
\email{ahaber@physics.wustl.edu}
\affiliation{Physics Department, Washington University in Saint Louis, 63130 Saint Louis, MO, USA}

\author{Ziyuan Zhang\,\orcidlink{0000-0003-4795-0882}}
\email{ziyuan.z@wustl.edu}
\affiliation{Physics Department, Washington University in Saint Louis, 63130 Saint Louis, MO, USA}
\affiliation{McDonnell Center for the Space Sciences, Washington University in Saint Louis, Saint Louis, MO 63130, USA}

\date{12 May 2024}   % Add date when submitting to arXiv

\begin{abstract}
We analyze the isospin equilibration properties of neutrinoless nuclear ($npe$) matter in the temperature and density range that is relevant to neutron star mergers. Our analysis incorporates neutrino-transparency corrections
to the isospin (``beta'') equilibrium condition which become
noticeable at $T\gtrsim 1\,\MeV$. We find that 
the isospin  relaxation rate rises rapidly as temperature rises, and at $T\approx 5\,\MeV$ it is comparable to the timescale of the density oscillations that occur immediately after the merger. This produces a resonant peak in the bulk viscosity at $T\approx 5\,\MeV$, which causes density oscillations to be damped on the timescale of the merger. Our calculations suggest that isospin relaxation dynamics may also be relevant when neutrinos are treated more accurately via neutrino transport schemes.

%We conclude that there is good reason to include isospin  relaxation dynamics in merger simulations. 
\end{abstract}

\maketitle

%%%%%%%%%%%%%%%%%%%%%%%%%%%%%%%%%%%
\section{Introduction}
\label{sec:intro}

Nuclear matter in neutron stars relaxes towards isospin  (``beta'')
equilibrium; its steady state has an equilibrium proton fraction $x_p^\text{eq}$
which is a function of
baryon density $n_B$ and temperature $T$.
Equilibrium is established by weak interactions which operate on a timescale that can range from microseconds to minutes. Astrophysical phenomena such as
density oscillations in neutron stars, which can be on a similar timescale, can therefore drive the system out of equilibrium, and the dynamics of the relaxation process may be relevant to our understanding of the astrophysics.

In this analysis we focus on the astrophysical conditions found in the central region of neutron star mergers, where homogeneous nuclear matter
at densities from one to several times nuclear saturation density $\nsat$ and temperatures up to
80\,MeV undergoes compression and rarefaction on millisecond timescales \cite{Baiotti:2016qnr,Alford:2017rxf}.
The purpose of this work is to argue that such oscillations are likely to
drive the system out of isospin  equilibrium \cite{Hammond:2021vtv}, and that
it is important for simulations to track its relaxation back to
chemical equilibrium, using the most accurate available Urca rate calculations, because
it may lead to physically important phenomena such as bulk viscosity.

A major complication is the role of neutrinos. There are two limits in which they can
be straightforwardly handled. The first is the ``neutrinoless'' limit where the contribution to flavor equilibration from neutrino absorption is negligible. This can be due to there being few neutrinos present and/or to their having a low absorption cross-section,
e.g.~at low temperatures. The second is
the ``neutrino-trapped'' regime where neutrinos have a short mean free path, e.g.~at high temperatures, and hence form a locally equilibrated Fermi gas.
In general, neutrinos in the central region of a merger are expected to lie between these two extremes,
and techniques are still being developed to handle their complex dynamics \cite{Perego:2019adq,Foucart:2022bth}.

In neutrino-trapped matter, isospin equilibration proceeds so fast that the relaxation dynamics can be neglected \cite{Alford:2019kdw,Alford:2021lpp,Espino:2023dei}, but previous work on neutrinoless matter found that isospin relaxation occurs on the same timescale as the fluid dynamics, giving rise to maximal bulk viscosity \cite{Alford:2019qtm,Alford:2021ogv, Most:2022yhe, Alford:2023uih}. This paper focuses on the neutrinoless regime, and provides
an improved treatment in various respects (see summary below) that aims to reinforce the earlier conclusions, namely that since isospin relaxation dynamics are important in one of the two extreme limits, there is good reason to do the best possible job of including them in more realistic neutrino transport schemes.  In the Conclusions we outline our
future plans towards this goal.

Given this motivation, we calculate physical quantities relevant for isospin  equilibration in neutrinoless homogeneous nuclear matter at densities above nuclear saturation density. We focus our calculation on the range $0<T<10\,\MeV$ because
neutrino trapping becomes stronger as temperatures rise beyond a few MeV \cite{Alford:2018lhf}. 
At these temperatures we can neglect positrons because at the densities we study the
electron chemical potential $\mu_e$ is in the $100\,\MeV$ range  so $T \ll \mu_e$.
We neglect muons because they introduce extra processes requiring a more sophisticated treatment.  Previous analyses have found that they do not make a large difference to the equilibration and relaxation rates \cite{Alford:2022ufz}, so we postpone their inclusion to future work. 
The quantities we calculate are the isospin   relaxation rate $\ga_I$ and the
the frequency-dependent bulk viscosity $\zeta$ and damping (sound attenuation) time for density oscillations.

Previous work on isospin relaxation in neutrinoless matter has assumed that the condition for isospin  equilibrium is
$\mu_n=\mu_p+\mu_e$ (e.g., 
\cite{Sawyer:1989dp,Haensel:1992zz,Haensel:2000vz,Jones:2001ya,Gusakov:2007px,Alford:2010gw,Schmitt:2017efp,Alford:2019qtm,Alford:2020lla,Alford:2023uih}). As we will now discuss, this is only valid at temperatures
below about $1\,\MeV$. At the temperatures attained in neutron star mergers there is a non-negligible correction to this condition. In this paper, we calculate the isospin relaxation rate
and bulk viscosity using the proper equilibrium condition.

In the infinite volume thermodynamic limit, equilibrium is established when the forward and backward rates of each process are equal (``detailed balance''), so the equilibrium condition is a simple equality involving chemical potentials. However, neutrino transparency is a finite-volume effect, arising when the mean free path of neutrinos is not much smaller than the size of the system. In a neutrinoless system neutrinos only occur in final states, never in initial states, so the system does not obey the principle of detailed balance. Equilibrium is attained when there is a balance between {\em different} processes: neutron decay and electron capture (as seen in Eqs.~\eqn{eq:dUrca} and \eqn{eq:mUrca} below).
In general, the equilibrium condition in neutrinoless matter
takes the form \cite{Alford:2021ogv, Alford:2018lhf}
\begin{equation}\label{eq:muIeq}
    \mu_n = \mu_p + \mu_e - \muIeq(n_B,T) \ ,
\end{equation}
where there is an isospin chemical potential $\muIeq$ whose value is determined by the requirement that the net rate of isospin creation is zero, i.e., the neutron decay and electron capture rates \eqn{eq:dU_plus_mU} are equal,
\begin{equation}
    \Ga_I \equiv \Gamma^\text{net}_{n\to p}  = 0 \ .
\end{equation}

At sufficiently low temperatures, $T\ll 1\,$MeV, the correction $\muIeq$ in Eq.~\eqn{eq:muIeq} becomes negligible because  the Fermi surface approximation is valid:
the Fermi-Dirac distribution exponentially suppresses contributions from particles away from their Fermi surfaces and the energy carried away by neutrinos is negligible (of order $T$). If neutrinos are kinematically negligible then neutron decay and electron capture are effectively the time-reverse of each other,
$n \leftrightharpoons p\ e^-$, and the isospin  equilibrium condition can be obtained from applying detailed balance to that process, yielding $\mu_n=\mu_p+\mu_e$, i.e., $\muIeq=0$.
However, as the temperature rises above about 1\,MeV, $\muIeq$ becomes non-negligible, and can reach values as large as  20~MeV (see Fig.~\ref{fig:muIeq} and Refs.~\cite{Alford:2021ogv, Alford:2018lhf}). 

In Sec.~\ref{sec:isospin -equilibration} we derive expressions for the
isospin   relaxation rate and the bulk viscosity in terms of the
microscopic isospin equilibration rate $\Ga_I$, i.e., the rate of Urca processes.
In Sec.~\ref{sec:nuclear_matter} we describe the relativistic mean field theories (RMFTs)
that we use to model nuclear matter. In Sec.~\ref{sec:Urca} we summarize the
calculation of the Urca rates and in Sec.~\ref{sec:results} we present our results.

We use natural units where $\hbar=c=k_B=1$.

\section{Isospin  equilibration of nuclear matter}
\label{sec:isospin -equilibration}

We now derive expressions for the isospin   relaxation rate $\ga_I$ and the bulk viscosity $\zeta$ of neutrinoless $npe$ matter at arbitrary temperature.
We will assume that matter always remains locally electrically neutral, so all calculations are performed at constant charge density $n_Q=0$.
The derivation in this section is applicable to both isothermal
and adiabatic density oscillations, by taking the derivatives
at constant temperature or constant entropy per baryon respectively.
In the presentation below we will not explicitly show the dependence on $T$ or $s/n_B$.

%The results presented in Sec.~\ref{sec:results} are calculated in  the isothermal regime.  We expect that results for the adiabatic regime will be similar \cite{Alford:2018lhf}: this is discussed further in Sec.~\ref{sec:conclusions}.

\subsection{Isospin  relaxation}
\label{sec:isospin -relaxation}

In neutrinoless nuclear $npe(\mu)$ matter, 
in addition to the conserved baryon number $B$
and conserved electric charge $Q$ 
there is another ``briefly-conserved'' charge, isospin $I$. The relevant charge densities and chemical potentials can be related to the net densities
and chemical potentials of neutrons ($n$), protons ($p$), and electrons ($e$),
\beq
\ba{rcr@{\,}c@{\,}r@{\,}c@{\,}r}
   n_B &=& n_p &+& n_n \ , \\[0.3ex]
   n_I &=& \half n_p &-& \half n_n  \ , \\[0.3ex]
   n_Q &=& n_p &&&-& n_e  \ , \\[1ex]
 \mu_B &=& \half\mu_p &+& \half\mu_n &+& \half\mu_e  \ , \\[0.3ex]
 \mu_I &=&  \mu_p &-& \mu_n &+& \mu_e  \ , \\[0.3ex]
 \mu_Q &=& &&&-& \mu_e \ .
 \ea
\label{eq:BIQ-def}
\eeq
The expressions above do not include a chemical potential for lepton number because the neutrinos are far from thermal equilibrium so there is no associated chemical potential.

It will be convenient to define the
isospin fraction $x_I$, which is simply related to the proton fraction 
\mbox{($x_p\equiv n_p/n_B$)},
\beq
x_I \equiv \dfrac{n_I}{n_B} = x_p - \half \ .
\label{eq:x-def}
\eeq
For many purposes, we can treat $x_I$ as
equivalent to $x_p$, since  $\p/\p x_I$ is the same as $\p/\p x_p$, and a derivative
at constant $x_I$ is also a derivative at constant $x_p$.

As noted in Sec.~\ref{sec:intro}, on strong-interaction ($\approx 10^{-23}$\,s)  timescales all three charges are conserved,
but on longer timescales the weak interactions break isospin,
so $\mu_I$ and $x_I$ relax to their
$\beta$ equilibrated values $\muIeq$ and  $x^\text{eq}_I$.
To analyze this equilibration process, where the system has been driven out of equilibrium by a density oscillation, it is natural to work in terms of $n_B$ and $x_I$, since
baryon density is the quantity that tracks the density oscillation, and the isospin fraction tracks relaxation to equilibrium. 

Equilibration of isospin is governed by the rate equation
\beq
\dfrac{d n_I}{dt} = \dfrac{n_I}{n_B}\dfrac{dn_B}{dt} 
  + \Gamma_I(n_B,x_I)  \ .
   \label{eq:Gamma-def}
\eeq
The first term on the right side tells us that if isospin were conserved then compression would change the isospin density by the same fraction as it changes baryon density. In the second term $\Gamma_I$ is the {\em isospin production rate}, i.e., the net rate per unit volume at which isospin increases, or equivalently the net rate at which neutrons are converted to protons. In Sec.~\ref{sec:Urca} we describe how it can be calculated from
the Fermi theory of weak interactions by integrating the net $n\to p$ rate over the Fermi-Dirac distributions of protons, neutrons, and electrons.

Using the definition \eqn{eq:x-def}, the rate equation
\eqn{eq:Gamma-def} becomes
\beq
\dfrac{d x_I}{dt} = \dfrac{1}{n_B} \Gamma_I(n_B,x_I)  \ ,
   \label{eq:xI-rate-general}
\eeq
and the equilibrium isospin fraction $x_I^\text{eq}(n_B)$ is defined by
\beq
\Gamma_I\bigl(n_B,x_I^\text{eq}(n_B,T)\bigr)=0 \ .
\label{eq:xI-eq-def}
\eeq
If $x_I$ is above its equilibrium value then there are too many protons, so the rate of $p\to n$ becomes larger than $n\to p$; $\Gamma_I$ should then become negative, driving $x_I$ back down towards its equilibrium value.
To obtain physically relevant quantities such as the
isospin relaxation rate and bulk viscosity we consider a generic small departure from equilibrium,
\beq
\Gamma_I(\bar n_B \!+\! \De n_B, \bar x_I \!+\! \De x_I) 
= \deriv{\Ga_I}{n_B}{x_I} \De n_B + \deriv{\Ga_I}{x_I}{n_B} \De x_I \, ,
\label{eq:Gamma-expansion}
\eeq 
where $\bar x_I \equiv x_I^\text{eq}(n_B)$.
Using this in the rate equation \eqn{eq:xI-rate-general}
we obtain the rate equation for the isospin fraction 
\beq
 \dfrac{d x_I}{dt} = -\ga_I \De x_I + \ga_B \dfrac{\De n_B}{\bar n_B} \, ,
\label{eq:xI-rate-linear}
\eeq
where
\beq
\ba{rl}
   \ga_I\equiv -\dfrac{1}{\bar n_B} \deriv{\Ga_I}{x_I}{n_B}\, , \\[3ex]
   \ga_B\equiv \deriv{\Ga_I}{n_B}{x_I} \, , 
\ea
\label{eq:gamma-def}
\eeq
with both the derivatives evaluated at $n_B=\bar n_B$,
\mbox{$x_I=\bar x_I$}. According to \eqn{eq:xI-rate-linear} a small deviation of $x_I$ from equilibrium (with no change in $n_B$) would evolve as
\mbox{$\dot x_I = - \ga_I (x_I - x_I^\text{eq})$}, so we identify
$\ga_I$ as the {\em isospin relaxation rate},
which we expect to be positive.

The other rate factor, $\ga_B$, tells us how quickly equilibrium is restored in response to a change in density at fixed isospin fraction. In previous treatments (e.g.~\cite{Sawyer:1989dp,Haensel:1992zz,Haensel:2000vz,Jones:2001ya,Gusakov:2007px,Alford:2010gw,Schmitt:2017efp,Alford:2020lla}) this was simply related to $\ga_I$ (see \eqn{eq:cold-gammaB}) because it was assumed that $\beta$ equilibrium corresponds to $\mu_I=0$; however, as noted in Sec.~\ref{sec:intro}, this is no longer true if the neutrinos are out of thermal equilibrium at $T\gtrsim 1\,\MeV$.

\subsection{Bulk viscosity}
\label{sec:bulk-viscosity}

To see how isospin  equilibration leads to bulk viscous damping, we consider
a fluid element of nuclear matter, with pressure $p$, that is driven
out of equilibrium by a small-amplitude density oscillation,
\beq
\ba{rcl}
n_B(t) &=&  \nbar + {\rm Re}( \de n_B\,  e^{i\om t})\ , \\[1ex]
x_I(t) &=& \bar x_I + {\rm Re}(\de x_I\, e^{i\om t}) \ , \\[1ex]
p(t) &=& \bar{p} + {\rm Re}(\de p\, e^{i\om t}) \ .
\ea
\label{eq:dV}
\eeq
We assume that the oscillation occurs around equilibrium, so
\beq
\bar x_I = x_I^\text{eq}(\bar n_B,T) \ .
\eeq
We adopt the phase convention that the baryon density amplitude $\de n_B \ll \bar n_B$ is real.
Bulk viscous dissipation arises from
a phase lag between pressure and density. The rate of energy dissipation for the small-amplitude oscillation \eqn{eq:dV} is 
obtained from the $pdV$ work done by the oscillation in one cycle. Rewriting $p dV$ as $p\,(dV/dt)\,dt$ the rate of energy dissipation per unit volume is
\beq
W = -\frac{1}{\tau \bar V}\int_0^\tau p(t)\frac{dV}{dt}dt 
  = \frac{1}{2} \om  \,{\rm Im}(\de p)\,\frac{\de n_B}{\nbar}\ ,
\label{W-dP}
\eeq
where the period is $\tau=2\pi/\om$ and we used \eqn{eq:dV} and the relation $n_B=N/V$ for a fluid element containing $N$ baryons.

The hydrodynamic relation between bulk viscosity and
rate of energy dissipation per unit volume is \mbox{$W = \zeta (\nabla\cdot \vec v)^2$}, which for the small-amplitude oscillation \eqn{eq:dV} becomes (averaged over one oscillation period)
\beq
W = \dfrac{1}{2} \zeta \om^2 \dfrac{(\de n_B)^2}{\nbar^2} \ .
\label{W-fluidmech}
\eeq
Identifying \eqn{W-dP} with \eqn{W-fluidmech} we obtain the frequency-dependent
bulk viscosity
\beq
\zeta(\om)  =  \frac{ {\rm Im}(\de p)}{\de n_B} \frac{\nbar}{\om} \ .
\label{eq:zeta-Im-p}
\eeq
For $npe$ nuclear matter, where the phase lag of the pressure arises
from the equilibration of isospin, this becomes
\beq
\zeta(\om) = \frac{\nbar}{\om}\, \deriv{p}{x_I}{n_B} 
 \dfrac{{\rm Im}(\de x_I)}{\de n_B} \ .
\label{eq:zeta}
\eeq

We can now obtain the bulk viscosity of nuclear matter 
by analyzing the equilibration process in more detail,
which will allow us to calculate the phase lag.
Substituting the explicit form of the oscillations
\eqn{eq:dV} in to the rate equation \eqn{eq:xI-rate-linear} we find the relationship between the amplitudes
$\de x_I$ and $\de n_B$
\beq
i\om \de x_I =  -\ga_I \de x_I + \dfrac{\ga_B}{\bar n_B} \de n_B \, ,
\eeq
which can be rewritten
\beq
\frac{\de x_I}{\de n_B} = \dfrac{1}{\bar n_B}\dfrac{\ga_B}{\ga_I + i\om} \ .
 %= \ga \chi^x_n\,\dfrac{\ga - i\om}{\ga^2 + \om^2} \ .
 \label{eq:dxIdnB}
\eeq
For the bulk viscosity \eqn{eq:zeta} we take the imaginary part of \eqn{eq:dxIdnB},
\beq
\zeta = - \deriv{p}{x_I}{n_B} \dfrac{\ga_B}{\ga_I^2 + \om^2} \ ,
\label{eq:bulk-visc-pderiv}
\eeq
where $\ga_B$ and $\ga_I$ are defined in \eqn{eq:gamma-def}
This is the general expression for the bulk viscosity, valid even when temperature corrections shift the equilibrium away from its low temperature limit $\muIeq=0$.
In Appendix~\ref{sec:cold-beta-equilibrium} we take the low temperature limit
and show that previous calculations, which assumed $\muIeq=0$, agree with the general result in that limit.
In the low temperature limit \eqn{eq:zeta-cold} it is clear that the dependence of bulk viscosity on density and temperature features a resonant maximum when the relaxation rate $\ga_I(n_B,T)$ coincides with the angular frequency $\om$ of the density oscillation. This is less clear in the more general expression \eqn{eq:bulk-visc-pderiv}, but we have found that for typical equations of state $\muIeq$ varies slowly enough as a function of $n_B$ and $T$ so that $\ga_B(n_B,T)$ is still roughly proportional to $\ga_I(n_B,T)$ (the constant of proportionality is a slowly varying function of $n_B$ and $T$), so the resonant peak is still present.

The bulk viscosity \eqn{eq:bulk-visc-pderiv} can also be written as
\beq
\zeta = \zeta_0\, \dfrac{\ga_I^2}{\ga_I^2 + \om^2} \ ,
\label{eq:bulk-visc-factorized}
\eeq
where 
\beq
\zeta_0 = - \deriv{p}{x_I}{n_B}  \dfrac{\ga_B}{\ga_I^2} 
\label{eq:bulk-visc-static}
\eeq
is the static (zero frequency) limit of the bulk viscosity.
(Note that in the isothermal regime the derivative of the pressure can be rewritten as a derivative of $\mu_I$, see App.~\ref{sec:maxwell}).
From the static bulk viscosity and the isospin relaxation rate $\ga_I$, which are functions of $n_B$ and $T$, one can reconstruct the full frequency-dependent bulk viscosity as a function of density and temperature.

%%%%%%%%%%%%%%%%%%%%%%%%%%%%%%%%%%%%%%%%%%%%%%%%%%%
\section{Nuclear matter models}
\label{sec:nuclear_matter}

One of the most important features influencing the isospin relaxation rate of nuclear matter is the direct Urca threshold density, which separates the low-density range, where in the $T\to  0$ limit only modified Urca processes are allowed, from the high-density range where direct Urca processes are kinematically allowed (see Sec.~\ref{sec:Urca}). It is not known whether real-world nuclear matter has a direct Urca threshold in the relevant density range, so we perform calculations for  two relativistic mean-field theories, IUF \cite{Fattoyev:2010mx} and QMC-RMF3 \cite{Alford:2022bpp}. Both are consistent with current astrophysical and nuclear constraints. 
IUF has a direct Urca threshold at $4~n_0$ whereas QMC-RMF3 does not have a threshold in the range of densities found in neutron stars. 

At a given baryon density $n_B$, temperature $T$, and proton fraction $x_p$ we solve the RMFT mean field equations to obtain the values of the meson condensates, thermodynamic quantities such as the pressure and proton and neutron chemical potentials, and also the nucleon effective masses and energy shifts.
The dispersion relations for nucleons in the relativistic mean field models are then specified as
\begin{equation}
    E_n = \sqrt{{m_n^*}^2 + k_n^2} + U_n \ ,
\end{equation}
\begin{equation}
    E_p = \sqrt{{m_p^*}^2 + k_p^2} + U_p \ ,
\end{equation}
where $m^*_i$ are the effective masses, $k_i$ are the particle momenta and $U_i$ are the energy shifts.
In RMFTs the effective masses and energy shifts are functions of density, temperature, and proton fraction.  In the models that we use the protons and neutrons have the same effective mass, so the energy shifts play an important role by separating the neutron and proton energies and thereby opening up more phase space for the Urca processes.

%%%%%%%%%%%%%%%%%%%%%%%%%%%%%%%%%%%
\section{Urca Rates}
\label{sec:Urca}

\subsection{Overview of Urca processes}

The isospin equilibration rate $\Ga_I$ \eqn{eq:Gamma-def} is given by the neutron decay and electron capture processes, which are governed by the weak interaction. Depending on the number of spectators, these processes are called direct or modified Urca. The direct Urca processes in neutrino transparent matter are neutron decay and electron capture
\begin{align}\label{eq:dUrca}
    n&\rightarrow p + e^- + \bar{\nu} \, ,\\
    p + e^- &\rightarrow n + \nu \, .\nonumber
\end{align}

In most models of homogeneous nuclear matter the proton fraction $x_p$ increases monotonically with the density. This means that for some models there is a direct Urca threshold density, defined as the density below which,
in the limit $T\to 0$, the rate of the direct Urca process is exponentially suppressed.
Since at low temperature the participating neutrons, protons, and electrons are all on their Fermi surfaces, the criterion for direct Urca to proceed is
$k_{Fn} \leqslant k_{Fp} + k_{Fe}$ where $k_{Fi}$ are the Fermi momenta for the particles.  In $npe$ matter this threshold is $x_p \geqslant 0.11$ since neutrality requires $k_{Fp}=k_{Fe}$. Below the threshold density, $k_{Fn} > k_{Fp} + k_{Fe}$, so momentum conservation forbids the direct Urca processes for particles on their Fermi surfaces. In order for momentum conservation to be satisfied, some particles would need to be away from their Fermi surfaces, but this
population is exponentially suppressed by their Fermi-Dirac distributions.
Consequently, the direct Urca rate is suppressed as $\exp(-\mathcal{E}/T)$ where $\mathcal{E}$ is an energy deficit that
goes to zero as the density reaches the dUrca threshold from below.
At densities below the direct Urca threshold (and temperatures well below the energy deficit for the given density) we expect the Urca rates to be dominated by the modified Urca process,
\begin{align}\label{eq:mUrca}
    N + n &\rightarrow N + p + e^- + \bar{\nu} \, ,\\
    N + p + e^- &\rightarrow N + n + \nu\nonumber \, ,
\end{align}
where $N$ is a spectator nucleon that can scatter from one part of its Fermi surface to another, injecting momentum via a virtual pion in to the accompanying direct Urca process.

We now discuss the calculation of the direct Urca (``dU'') and modified Urca (``mU'') rates, which establish $\beta$ equilibrium (see Sec.~\ref{sec:isospin -equilibration}),
\begin{align}\label{eq:dU_plus_mU}
    \Gamma_\text{nd} &= \Gamma_\text{dU,nd} + \Gamma_\text{mU,nd} \, , \\
    \Gamma_\text{ec} &= \Gamma_\text{dU,ec} + \Gamma_\text{mU,ec} \, .
\end{align}

In this paper, we will calculate the direct Urca rates by integrating over the full phase space, including Fermi-Dirac-suppressed contributions that become non-negligible when the temperature rises to the MeV range \cite{Alford:2018lhf}. We will calculate modified Urca rates in the Fermi surface approximation (in which participating nucleons and electrons are assumed to be on their Fermi surfaces) since modified Urca always has a non-suppressed contribution from such particles.

%The rates have a strong temperature dependence. Under Fermi surface approximation, at a given density, direct Urca increases as $T^5$ and modified Urca increases as $T^7$ [cite].

\subsection{Direct Urca rates}
\label{sec:dUrca}

The direct Urca neutron decay and electron captures rates are \cite{Yakovlev:2000jp}
\begin{align}
    \Gamma_{\text{nd}}=\int & \frac{d^3k_n}{(2\pi)^3}\frac{d^3k_p}{(2\pi)^3}\frac{d^3k_e}{(2\pi)^3}\frac{d^3k_\nu}{(2\pi)^3} \nonumber\\
 &   f_n(1-f_p)(1-f_e)  \frac{\sum |M|^2}{(2E^*_n)(2E^*_p)(2E_e)(2E_{\nu})} \nonumber\\
  & (2\pi)^4\delta^{(4)}(k_n-k_p-k_e-k_\nu)\,,
    \label{eq:durca_nd_integral}
\end{align}
and 
\begin{align}
    \Gamma_{\text{ec}}=\int & \frac{d^3k_n}{(2\pi)^3}\frac{d^3k_p}{(2\pi)^3}\frac{d^3k_e}{(2\pi)^3}\frac{d^3k_\nu}{(2\pi)^3} \nonumber\\
    & (1-f_n) f_p f_e \, \frac{\sum |M|^2}{(2E^*_n)(2E^*_p)(2E_e)(2E_{\nu})} \nonumber \\
    & (2\pi)^4\delta^{(4)}(k_p + k_e - k_n - k_\nu)\,,
    \label{eq:Gamma-ec-def}
\end{align}
where $\Sigma|M|$ is the spin-summed matrix element, \mbox{$E_i^* = \sqrt{k_i^2+m_i^{*2}}$} are the effective nucleon dispersion relations, \mbox{$E_e=\sqrt{k_e^2+m_e^2}$} and \mbox{$E_\nu=k_\nu$} are the electron/neutrino dispersion relations, and $f_i$ are Fermi-Dirac distributions.
Direct Urca rates are often calculated in various approximations such as  using non-relativistic dispersion relations for the nucleons \cite{Haensel:2000vz,Alford:2018lhf,Alford:2019kdw,Alford:2020lla}, using vacuum masses for the nucleons \cite{Alford:2018lhf}, simplifying the matrix element \cite{Alford:2019kdw,Alford:2020lla}, or using the Fermi surface approximation \cite{Haensel:2000vz}. 

Our direct Urca rate calculations use the complete matrix element, relativistic dispersion relations for all participating particles, and integrate momenta over the entire phase space. For details see Appendix A of Ref.~\cite{Alford:2021ogv}.

\subsection{Modified Urca rates}
\label{sec:mUrca}

We use the standard expressions for the modified-Urca neutron decay and electron captures rates,
\begin{align}
    \Gamma_{mU,nd} &=\int \frac{d^3k_n}{(2\pi)^3} \frac{d^3k_p}{(2\pi)^3} \frac{d^3k_e}{(2\pi)^3} \frac{d^3k_{\nu}}{(2\pi)^3} \frac{d^3k_{N_1}}{(2\pi)^3} \frac{d^3k_{N_2}}{(2\pi)^3}\;\nonumber\\
    &(2\pi)^4 \delta^{(4)}(k_n+k_{N_1}-k_p-k_e-k_{\nu}-k_{N_2})\nonumber\\
    &f_n f_{N_1}(1-f_p)(1-f_e)(1-f_{N_2})\nonumber\\
    &\Bigl(s \frac{\sum |M|^2}{2^6 E^*_n E^*_p E_e E_{\nu} E^*_{N_1} E^*_{N_2}}\Bigr),
\end{align}
and
\begin{align}
    \Gamma_{mU,ec} &=\int \frac{d^3k_n}{(2\pi)^3} \frac{d^3k_p}{(2\pi)^3} \frac{d^3k_e}{(2\pi)^3} \frac{d^3k_{\nu}}{(2\pi)^3} \frac{d^3k_{N_1}}{(2\pi)^3} \frac{d^3k_{N_2}}{(2\pi)^3}\;\nonumber\\
    &(2\pi)^4 \delta^{(4)}(k_p+k_e+k_{N_1}-k_n-k_{\nu}-k_{N_2})\nonumber\\
    &f_p f_e f_{N_1}(1-f_n)(1-f_{N_2})\nonumber\\
    &\Bigl(s \frac{\sum |M|^2}{2^6 E^*_n E^*_p E_e E_{\nu} E^*_{N_1} E^*_{N_2}}\Bigr)\,,
\end{align}
where $s=1/2$ is a symmetry factor to account for identical particles. These are obtained using  the matrix element from
Ref.~\cite{Friman:1979ecl,Haensel:2001mw} and using the Fermi surface approximation to simplify the phase space integral.  For details see Appendix~B of \cite{Alford:2021ogv}.

%%%%%%%%%%%%%%%%%%%%%%%%%%%%%%%%%%%
\section{Results}
\label{sec:results}

We now present numerical results for the quantities
described in Sec.~\ref{sec:isospin -equilibration}.  We perform
all the computations in the isothermal regime: see Sec.~\ref{sec:conclusions} for further discussion of this assumption.

\subsection{Isospin chemical potential} 

To calculate linear-response isospin  equilibration properties such as the relaxation rate or bulk viscosity one needs to perturb around the equilibrium state. i.e., the state in which the net rate of isospin creation is zero.
As noted in Sec.~\ref{sec:intro},  at nonzero temperature this requires a nonzero isospin chemical potential, $\muIeq$ (Eq.~\eqn{eq:muIeq}). Its value  is negative because thermal corrections enhance electron capture more than neutron decay, so to restore  the balance between these rates we need a chemical potential that reduces the proton fraction: this suppresses electron capture by reducing the proton population and enhances neutron decay by adding more occupied neutron states.

\begin{figure*}[t]
\includegraphics[width=0.48\textwidth]{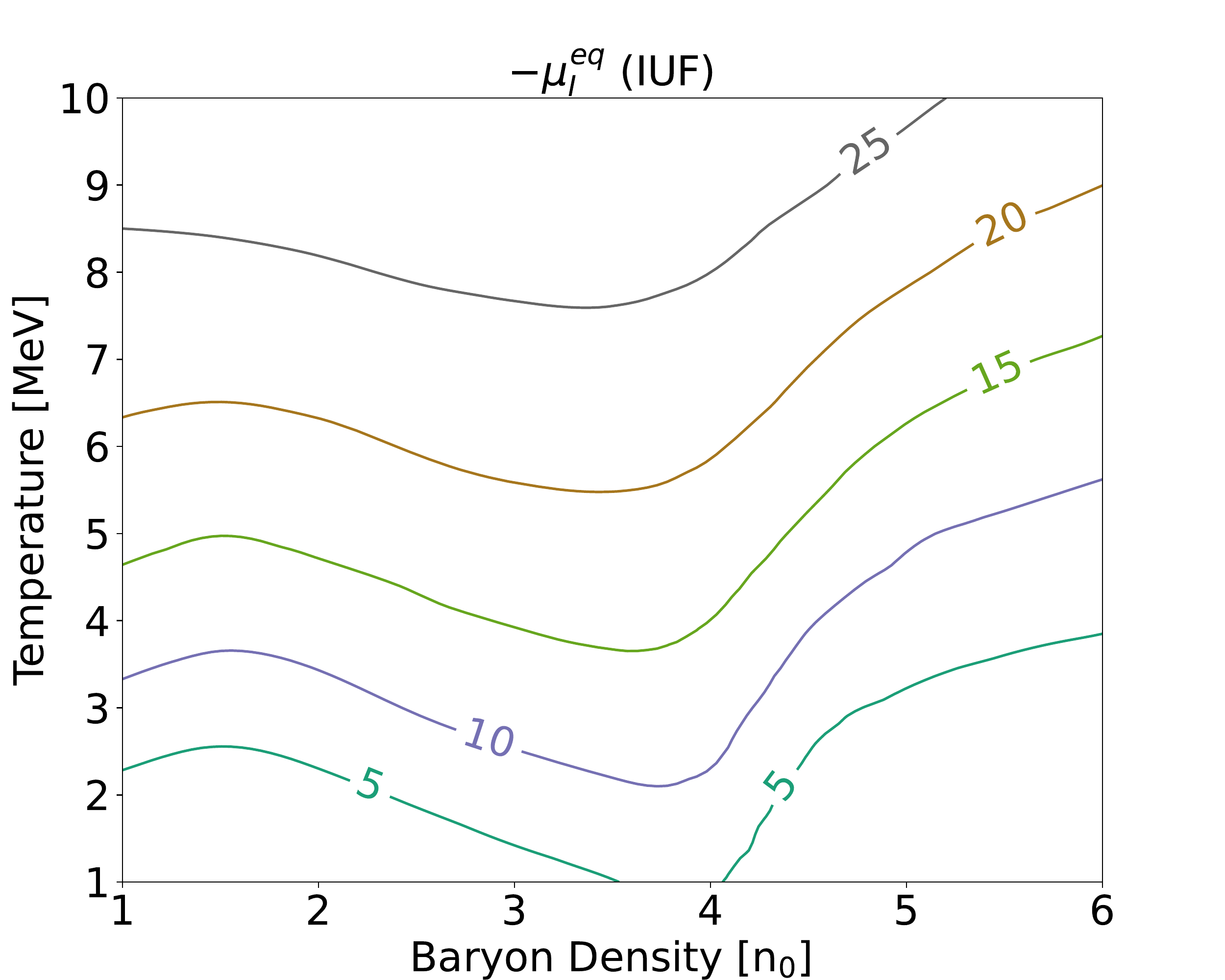}
\includegraphics[width=0.48\textwidth]{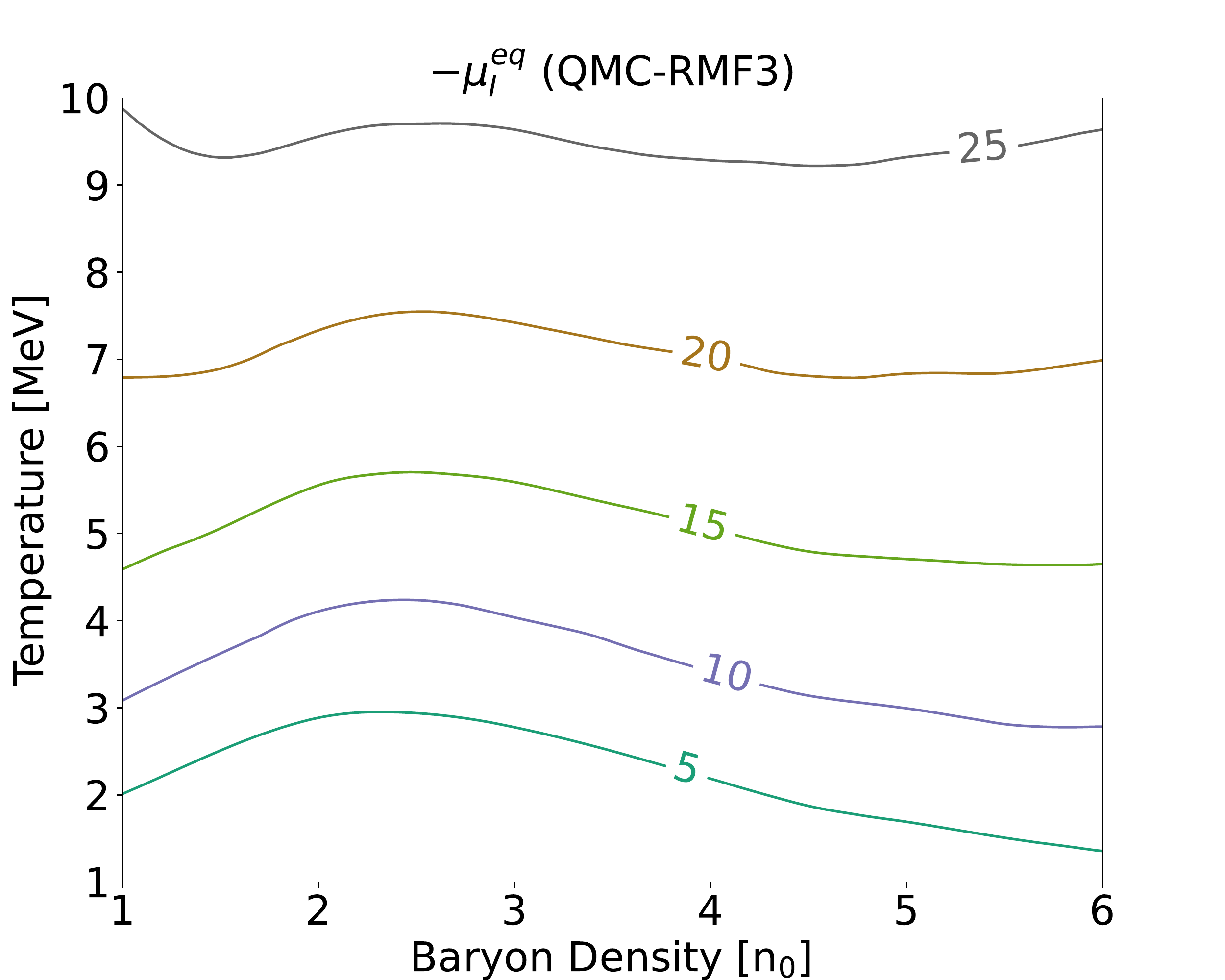}
\caption{
Density and temperature dependence of $-\muIeq$, the isospin chemical potential in isospin  equilibrium (Eq.~\eqn{eq:muIeq}), for the IUF equation of state (left panel) and QMC-RMF3 (right panel); $-\muIeq$ rises with temperature because it arises from thermal blurring of the
Fermi surfaces (see text). For IUF at low temperatures it is also influenced by the  direct Urca threshold at $n_B\approx 4\nsat$.
}
\label{fig:muIeq}
\end{figure*}

In Fig.~\ref{fig:muIeq} we show how $-\muIeq$ varies with density and temperature for our two exemplary equations of state, IUF and QMC-RMF3.
The plot shows contours labeled by their value of $-\muIeq$.
As discussed in Refs.~\cite{Alford:2018lhf,Alford:2021ogv} $\muIeq$
tends to zero as $T\to 0$ which is why it was neglected in previous treatments of bulk viscosity in neutron stars.  However, as the temperature rises above about 1\,MeV, it becomes non-negligible. 
Note that for IUF, which has a direct Urca threshold at $n_B\approx 4\nsat$,
$\muIeq$ is enhanced near the threshold density. This is because thermal blurring of the Fermi surfaces becomes more important close to threshold, and opens up phase space for the electron capture process more than it does for neutron decay \cite{Alford:2018lhf,Alford:2021ogv}; this imbalance is the reason for a nonzero $\muIeq$.

\subsection{Isospin relaxation rate}

\begin{figure*}[t]
\includegraphics[width=0.48\textwidth]{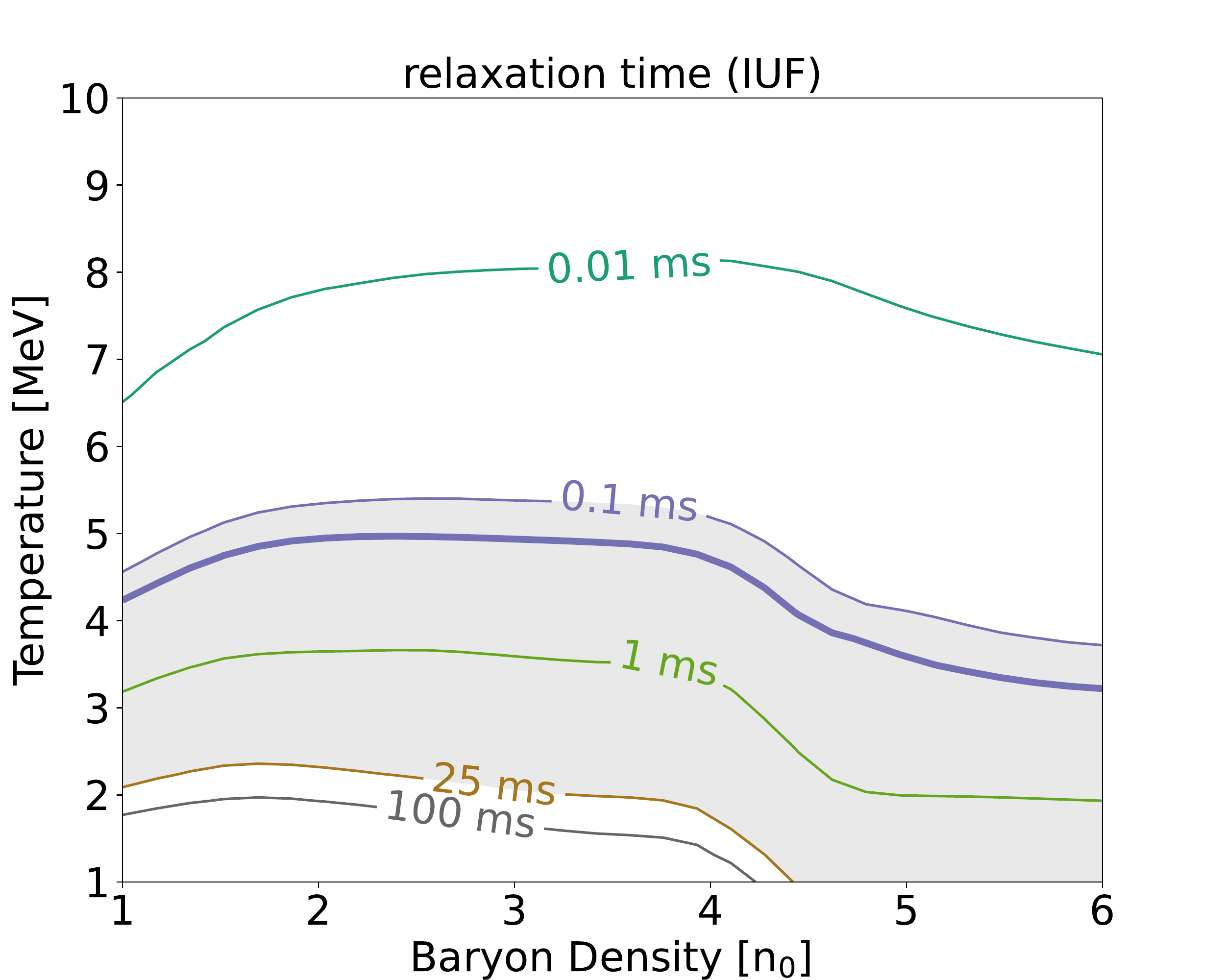}
\includegraphics[width=0.48\textwidth]{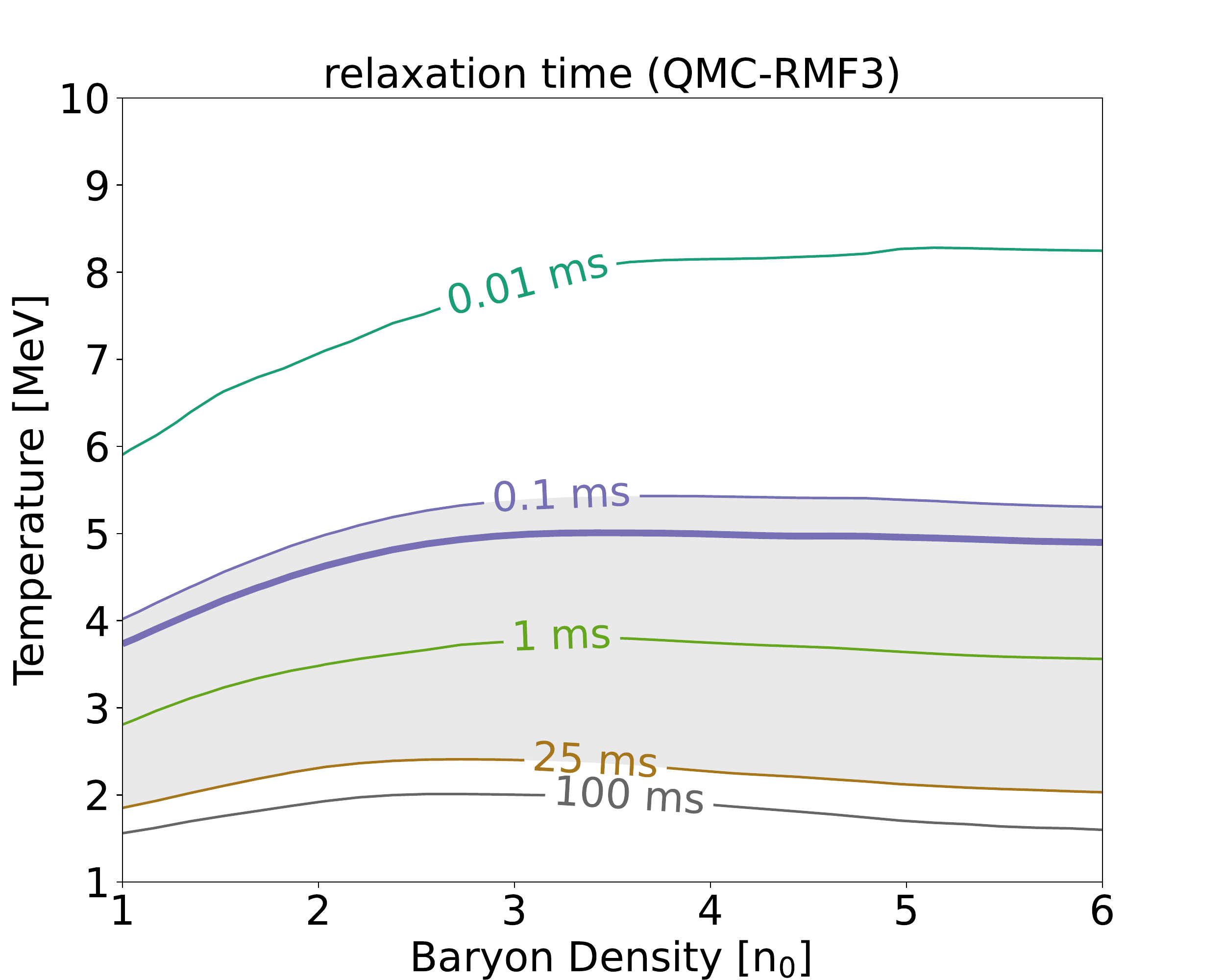}
\caption{
Density and temperature dependence of the isospin relaxation time
$\tau=1/\ga_I$ (Eq.~\eqn{eq:gamma-def}).  The shaded region
shows where
the relaxation time is between 0.1\,ms and 25\,ms, i.e. comparable to
the timescale of merger dynamics. 
The thick line shows the temperatures and densities where the bulk viscosity of a $1\,\kHz$ density oscillation would reach its maximum, i.e., where
$\ga_I=2\pi\times 1\,\kHz$ (Sec.~\ref{sec:bulk-viscosity}).
}
\label{fig:relax_time}
\end{figure*}

Fig.~\ref{fig:relax_time} shows how the isospin   relaxation time $\tau=1/\gamma_I$ \eqn{eq:gamma-def} depends on density and temperature for our two reference equations of state, IUF (left) and QMC-RMF3 (right).  We have shaded  the range of density and temperature where relaxation occurs on the timescale relevant for mergers, 0.1\,ms to 25\,ms. The thick contour shows where $\ga_I=2\pi\!\times\!1\,\kHz$, which as we will see below is where the bulk viscosity reaches a resonant maximum for a $1\,\kHz$ oscillation.
If there is material in a merger that lies in this density and temperature range and obeys our assumptions (such as neutrino transparency), then the relaxation of its proton fraction occurs on a timescale that is comparable to that of the merger dynamics, indicating that the relaxation process should be included in simulations.

In regions where the equilibration time is much smaller than $0.1$\,ms 
equilibration happens so fast that one could use the approximation that the matter is always in isospin  equilibrium. In regions where the equilibration time is much longer than $20$\,ms the equilibration process is slow, and the proton fraction of each fluid element could be approximated as being constant.
Previous simulations of neutron star mergers have either investigated these extreme cases  of instantaneous equilibration, or frozen composition \cite{Celora:2022nbp} or only partly implemented the low density and low-temperature approximation to the Urca rates studied here \cite{Ruffert:1996by,Most:2019kfe, Foucart:2022bth,Zappa:2022rpd,Musolino:2023pao}. 
 
Recently a first attempt has been made  \cite{Most:2022yhe} to include both direct and modified Urca processes, calculated in the Fermi surface approximation, in the simulation. It was found that Urca processes affect the
proton fraction of the fluid elements on the timescale of the merger dynamics.

For the IUF equation of state there is a noticeable feature in the relaxation time plot: relaxation becomes faster when the density reaches around $4 n_0$. This is because IUF has a direct Urca threshold at this density.
At densities above this threshold more phase space opens up and Urca rates, at a given temperature, are much faster than they are at densities below the threshold. 
%Therefore, above the threshold the relaxation rate a lower temperature is required for the Urca processes to happen on the relevant timescale. 
Below the threshold density, the relaxation time is comparable to the merger timescale at temperatures of order 2 to 4\,\MeV. Above the threshold density, the relaxation time is comparable to the merger timescale at temperatures below 2\,\MeV.

For QMC-RMF3, the relaxation time depends only weakly on density. This is because there is no direct Urca threshold. Across the whole density range that we studied $\beta$ equilibrium is determined by a balance between neutron decay (dominated by modified Urca) and electron capture (dominated by direct Urca) \cite{Alford:2018lhf,Alford:2021ogv}. Consequently, the region where the relaxation time is comparable to the merger timescale extends across
the whole density range that we computed, for temperatures in the \MeV-range.

Assuming that our reference equations of state (EoSs) are representative,  Fig.~\ref{fig:relax_time} implies that if neutrinoless homogeneous matter is present in mergers then
regions at $T\lesssim 5\,\MeV$ will likely be driven out of isospin  equilibrium  and regions at $T\approx 2$ to $5\,\MeV$ (the exact range depending on the EoS) will equilibrate on the timescale of the merger.

\subsection{Bulk viscosity and damping time}

\begin{figure*}[t]
\includegraphics[width=0.48\textwidth]{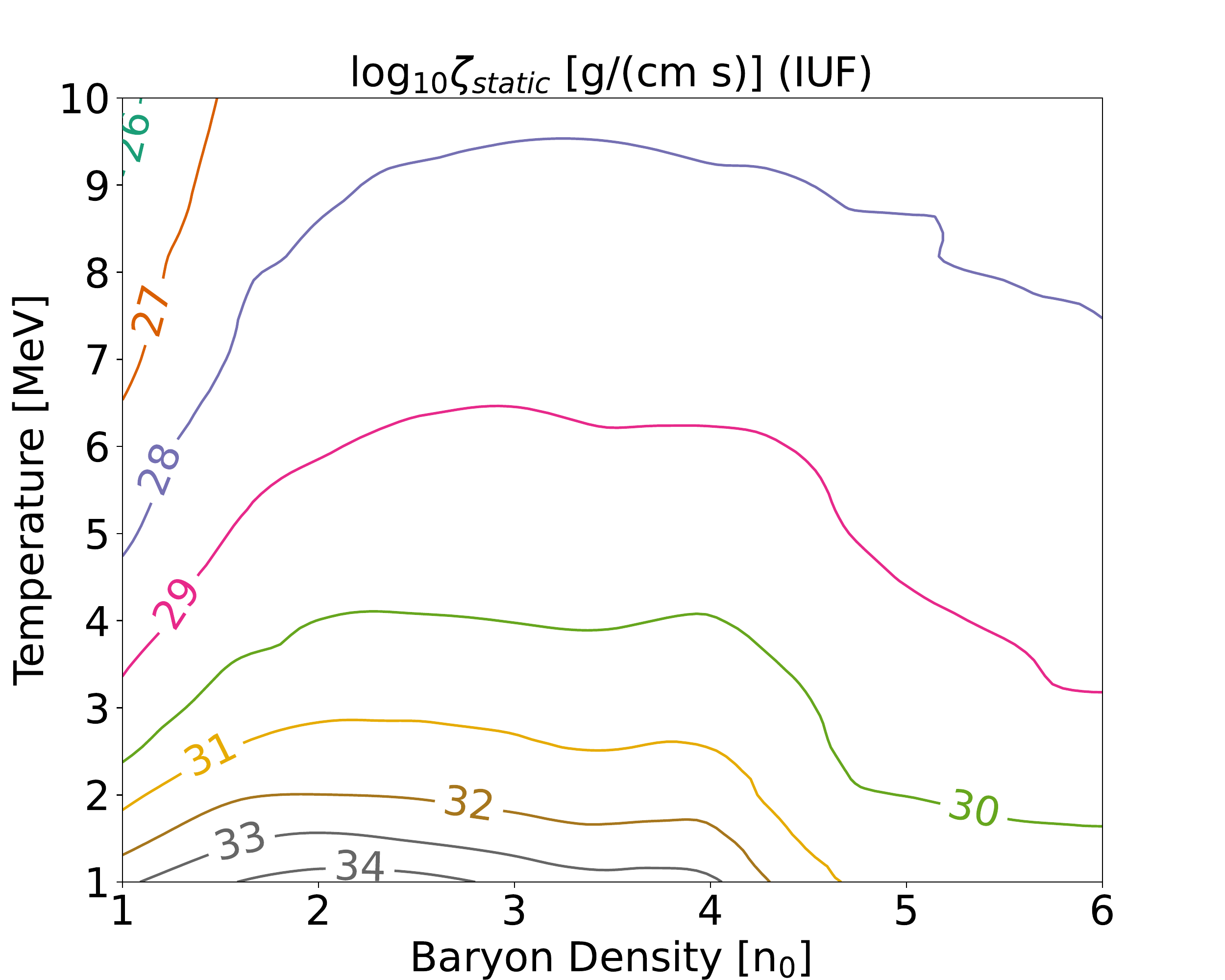}
\includegraphics[width=0.48\textwidth]{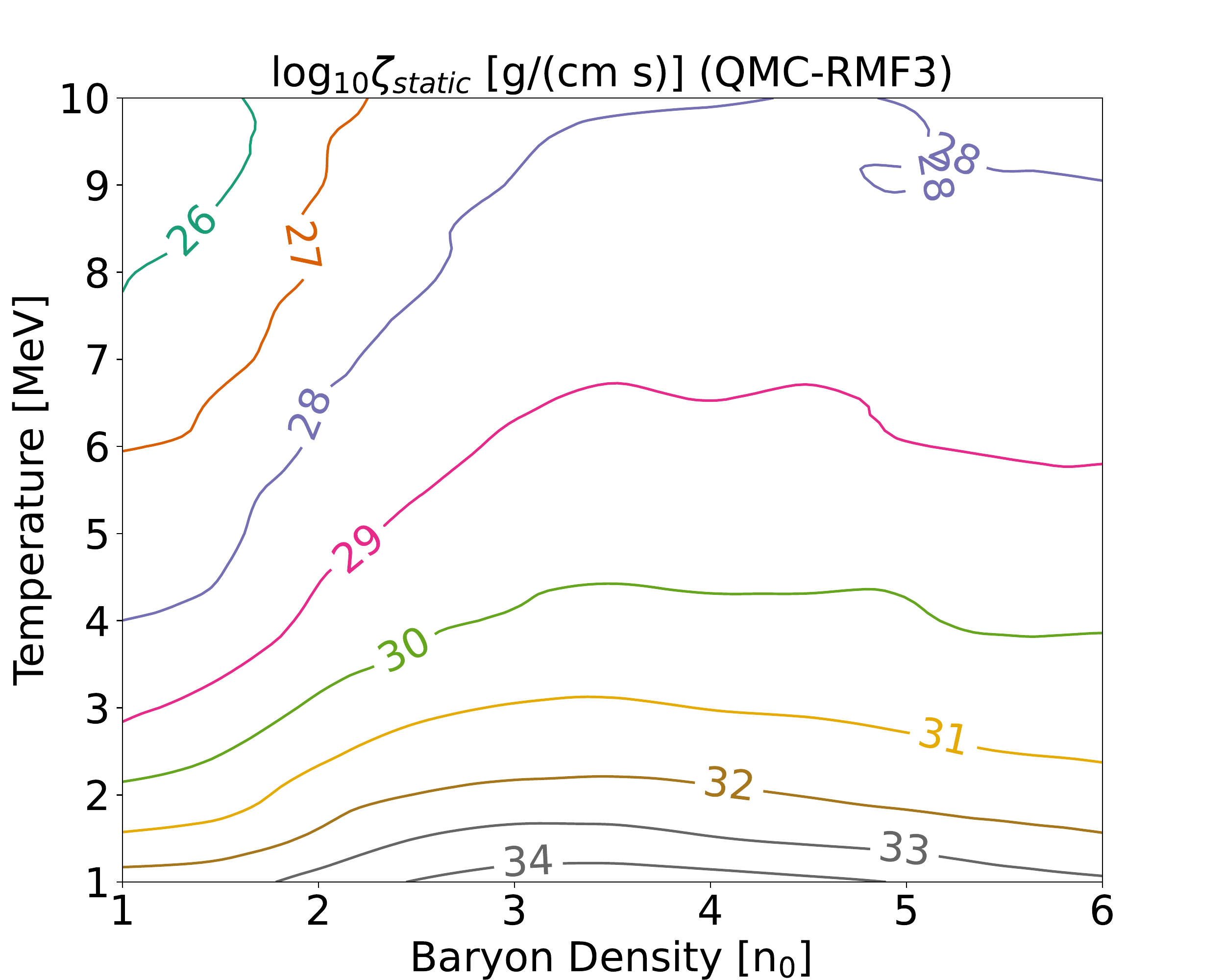}
\caption{
Static (zero-frequency) bulk viscosity \eqn{eq:bulk-visc-static} for the IUF (left) and QMC-RMF3 (right) equations of state. The static bulk viscosity drops as temperature rises because it is inversely proportional to the relaxation rate.
}
\label{fig:bv_static}
\end{figure*}

In Fig.~\ref{fig:bv_static} we show the density and temperature dependence of the static (zero-frequency) bulk viscosity $\zeta_0$ \eqn{eq:bulk-visc-static}.
The plot shows contours of $\log_{10}(\zeta_0)$ where $\zeta_0$ is in cgs units ($\gram\,\cm^{-1}\,\sec^{-1}$).  
The static bulk viscosity depends inversely on the relaxation rate, so it drops as the temperature rises, and at low temperatures we also see the effects of the direct Urca threshold at $n_B\approx 4\nsat$, as in Fig.~\ref{fig:relax_time}.

\begin{figure*}[t]
\includegraphics[width=0.48\textwidth]{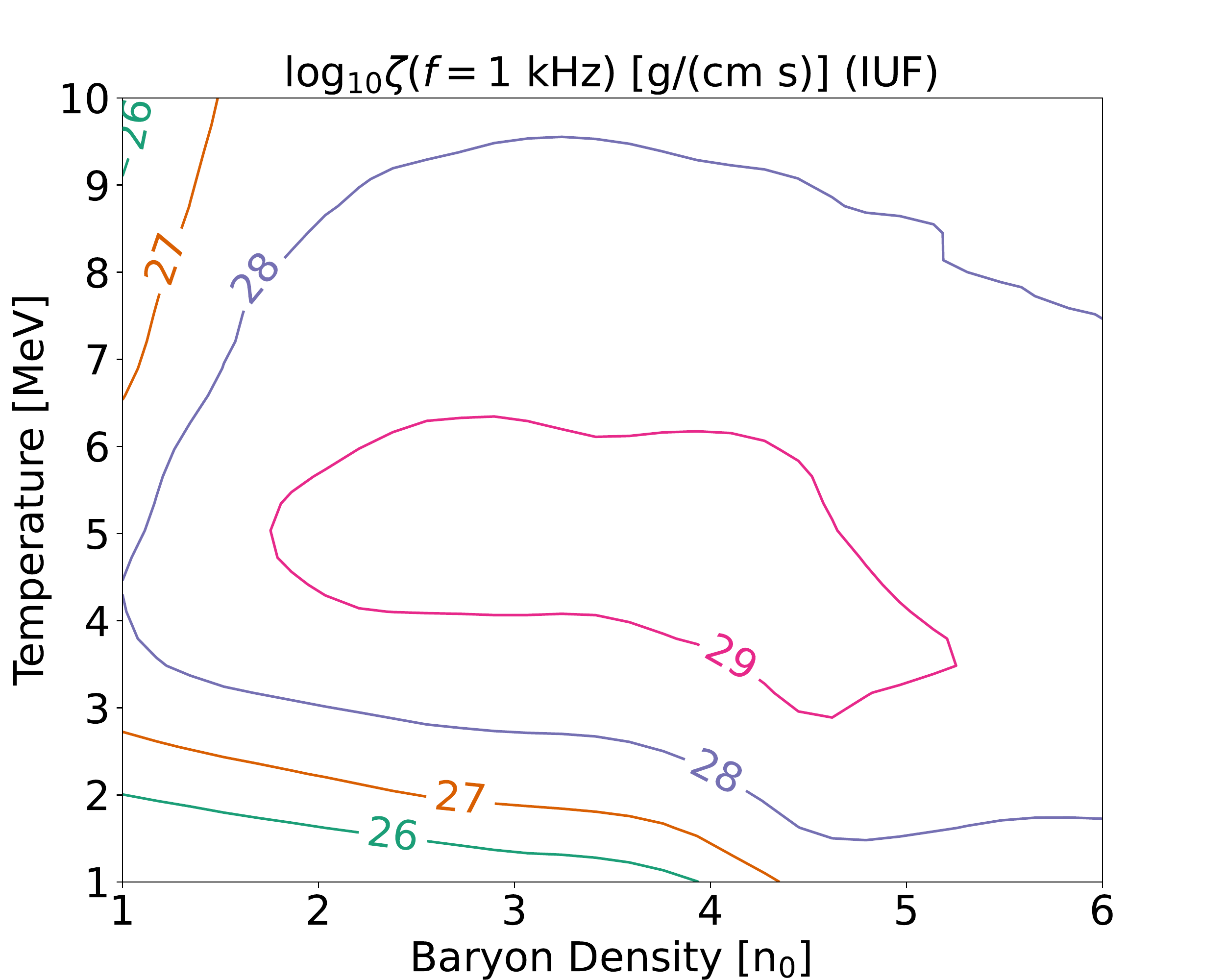}
\includegraphics[width=0.48\textwidth]{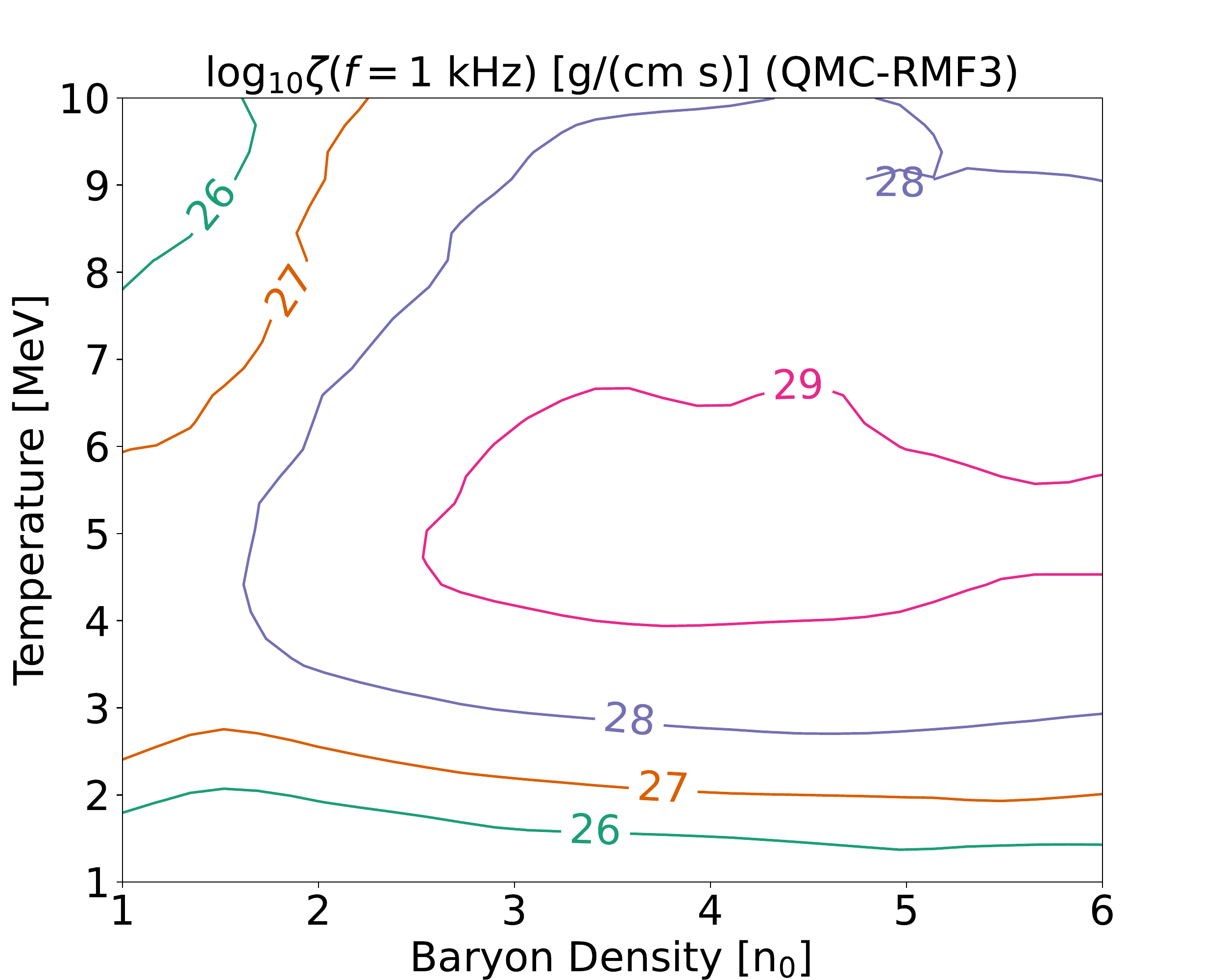}
\caption{ 
Bulk viscosity at frequency $f=1\,\kHz$ for the IUF (left) and QMC-RMF3 (right) equations of state. In both cases the resonant peak occurs where the relaxation rate passes through $\om=2\pi\times 1\,\kHz$, which occurs at $T\approx 5\,$MeV.
}
\label{fig:bv}
\end{figure*}

From the isospin relaxation rate and the static bulk viscosity one can reconstruct the frequency-dependent bulk viscosity (Eq.~\eqn{eq:bulk-visc-factorized}).  Since density oscillations in neutron star mergers typically have frequencies in the kHz range,
Fig.~\ref{fig:bv} shows the density and temperature dependence of 
bulk viscosity at angular frequency $\om=2\pi\!\times\!\,\kHz$.

As described in Sec.~\ref{sec:bulk-viscosity}, we expect the bulk viscosity to reach a resonant maximum when the isospin relaxation rate $\ga_I(n_B,T)$ coincides with the
angular frequency $\om$ of the density oscillation. The relaxation rate rises quickly with temperature since higher temperature opens up more phase space near the Fermi surfaces.  We therefore expect the bulk viscosity to achieve its maximum value at the temperature where $\ga_I(n_B,T)\approx \om$. From
Fig.~\ref{fig:relax_time} we see that for a 1\,\kHz\ density oscillation that
temperature is around 5\,\MeV.
This explains what we see in Fig.~\ref{fig:bv}: the contours  run roughly horizontally, with the bulk viscosity reaching a maximum at $T\approx 5\,\MeV$.
At lower temperatures the system equilibrates so slowly that isospin is almost conserved: the proton fraction remains constant, and the system has a low bulk viscosity.
At higher temperatures where $\ga_I \gg \om$ the system equilibrates so quickly that there is little phase lag between pressure and density, and the bulk viscosity tends towards its static value $\zeta_0$ (Eq.~\ref{eq:bulk-visc-static}).

For the IUF EoS one can see the effect of the direct Urca threshold at $n_B\approx 4\nsat$: below that density relaxation is slower (due to a lack of kinematically allowed phase space), and so a higher temperature is needed to bring the relaxation rate up to 1\,\kHz.

Comparing these to previous calculations that used nonrelativistic dispersion relations for the nucleons (e.g., Refs.~\cite{Alford:2020lla}, \cite{Alford:2019qtm}), our
resonance peaks are shifted to slightly higher temperatures for given densities. This is because in our models of nuclear matter the in-medium effective masses of the nucleons are much lower than the vacuum masses so the relativistic corrections are significant, decreasing the Urca rates and relaxation rates \cite{Alford:2021ogv,Alford:2022ufz}, which means higher temperatures are required to achieve resonance ($\ga_I=\om$).

\begin{figure*}[t]
\includegraphics[width=0.48\textwidth]{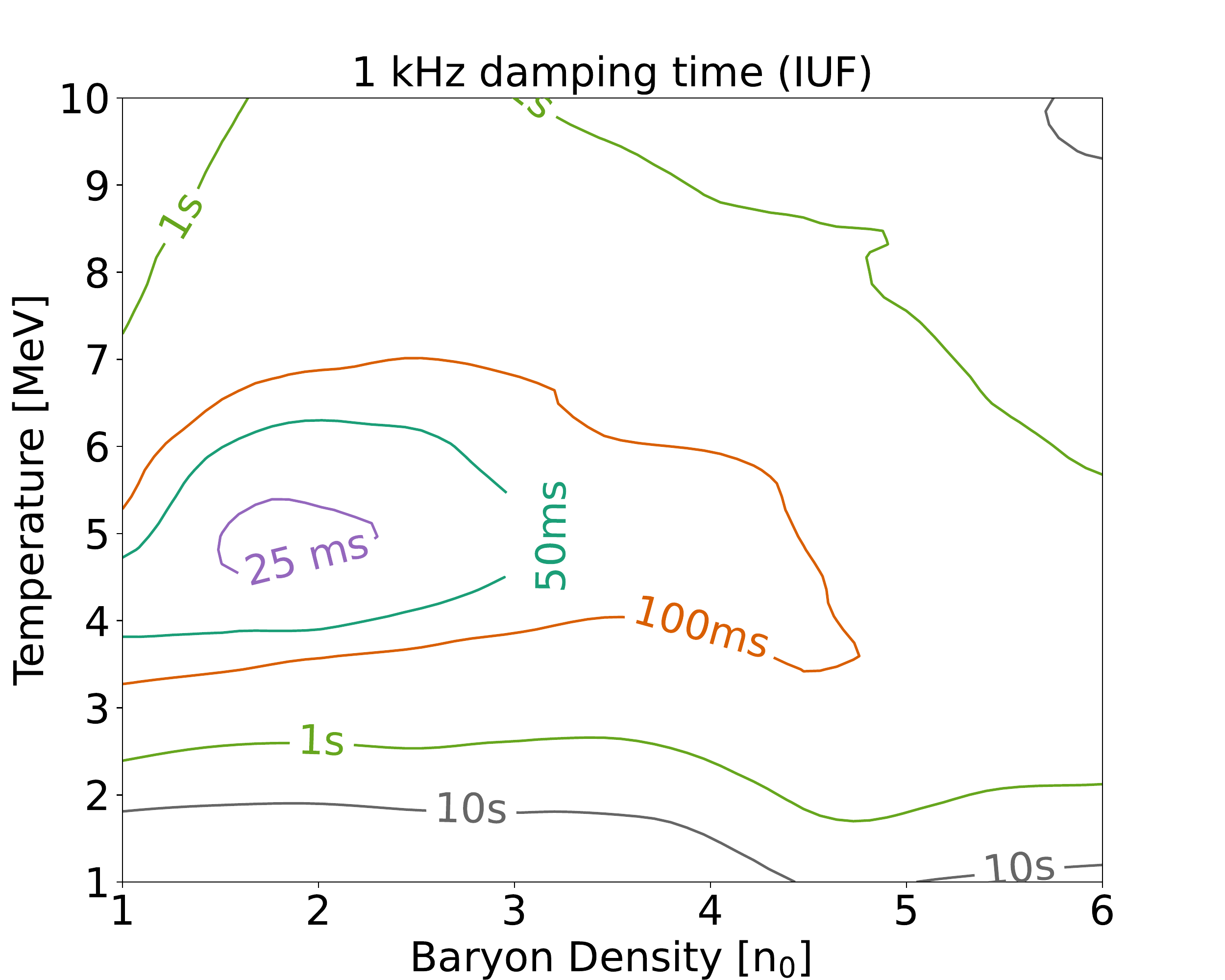}
\includegraphics[width=0.48\textwidth]{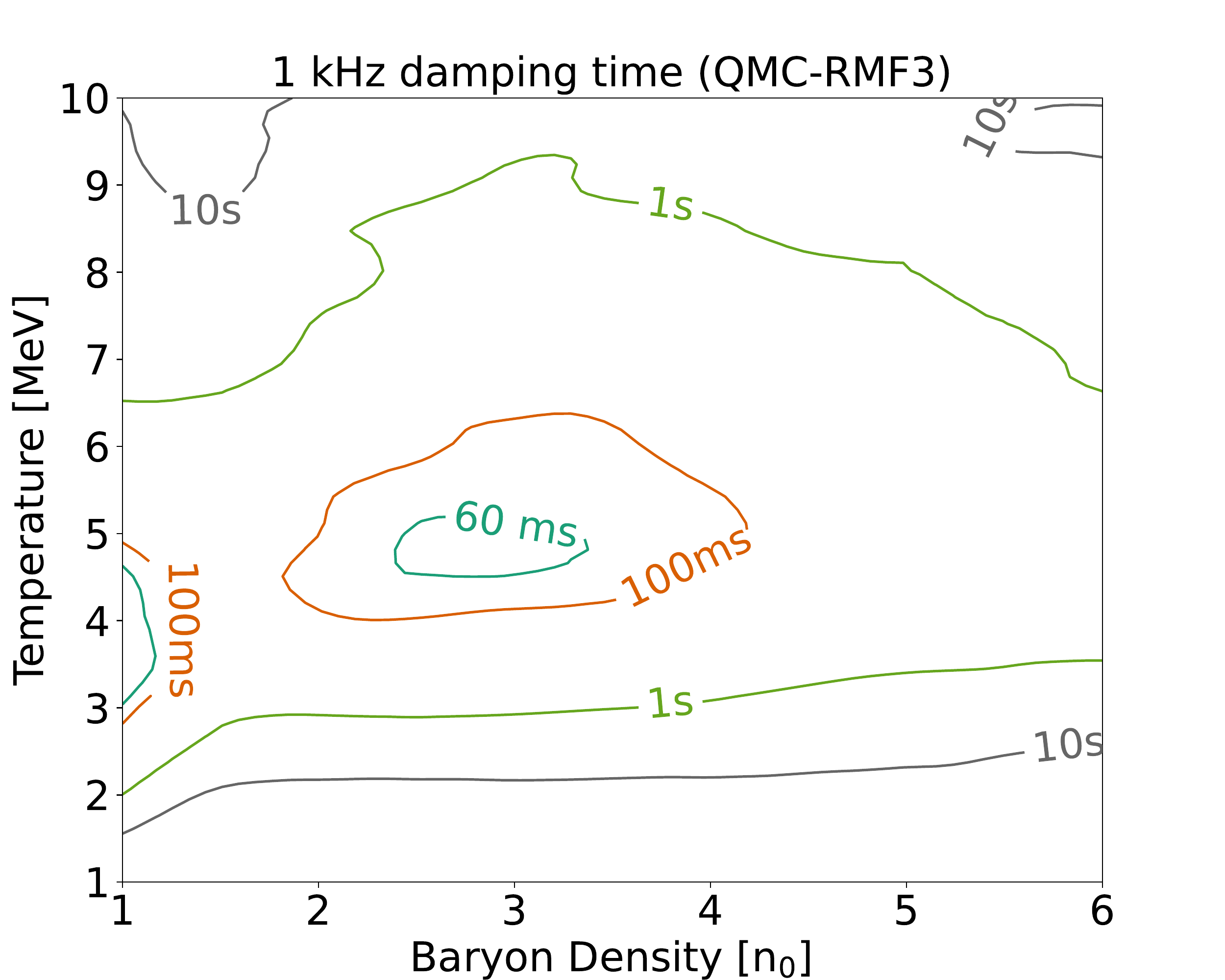}
\caption{ 
Damping time for density oscillations of frequency
1\,\kHz\ as a function of density and temperature, for the
IUF (left) and QMC-RMF3 (right) equations of state. }
\label{fig:damping_time}
\end{figure*}

One physical manifestation of bulk viscosity is the damping of density oscillations. The damping time for an oscillation of angular frequency $\om$ is \cite{Alford:2019qtm}
\beq
\tdamp = \dfrac{\ka^{-1}}{\om^2 \zeta(\om)} \, ,
\eeq
where the incompressibility is
\beq
\ka^{-1} \equiv n_B \dfrac{\p p}{\p n_B}\biggr|_{x_I,T} \, .
\eeq
Since this paper focuses on isothermal density oscillations we use the isothermal compressibility.

Fig.~\ref{fig:damping_time} shows how the damping time depends on density and temperature for IUF (left) and QMC-RMF3 (right). We expect that in density and temperature regions where the damping time is in the tens of milliseconds range, bulk viscosity will have a significant impact on density oscillations during the merger.

The key features of this plot are:\\
(1) the temperature dependence of the damping time is mainly determined by that of the bulk viscosity, so the damping time is shortest when the bulk viscosity (Fig.~\ref{fig:bv}) is largest, i.e. at $T\approx 5\,\MeV$;\\
(2) the density dependence of the damping time also roughly follows that of the bulk viscosity, but damping is slowed at high densities by the growth of the incompressibility: oscillations then store more energy and so take longer to decay; \\
(3) the shortest damping times are short enough so that bulk viscous damping is relevant on merger timescales.

These results are comparable to those obtained in Ref.~\cite{Alford:2022ufz}, which used different models for nuclear matter and used the low-temperature approximation for $\beta$ equilibrium, $\muIeq=0$.

The data 
for the plots shown in this section are available at \url{https://gitlab.com/ahaber/npe-bulkviscosity}. The code used to develop the QMC-RMF3 model and to solve the mean-field equations for both our models of nuclear matter can be found at \url{https://gitlab.com/ahaber}. The code for calculating Urca rates and isospin  equilibration properties will be made public as part of the MUSES framework  \url{https://musesframework.io/}.

\section{Conclusions}
\label{sec:conclusions}

We have analyzed the isospin   equilibration properties of neutrinoless nuclear ($npe$) matter in the temperature and density range that is relevant to neutron star mergers. Our analysis includes corrections to the isospin  equilibrium condition $\mu_n=\mu_p+\mu_e$ which arise at $T\gtrsim 1\,\MeV$.
We find that at temperatures of order 2 to $5\,\MeV$ the isospin  relaxation time, i.e., the timescale on which the proton fraction relaxes to its equilibrium value, is comparable to the timescale of the density oscillations that occur immediately after the merger. At lower temperatures, isospin  relaxes more slowly, and at higher temperatures, it relaxes faster. For a 1\,kHz density oscillation this leads to a resonant peak in the bulk viscosity at $T\approx 5\,\MeV$, when the relaxation rate matches the frequency. This causes damping of such a density oscillation on the timescale of the merger, providing evidence that when neutrinos are treated more rigorously it may still be important
to include isospin  relaxation dynamics in merger simulations. 

There are many directions in which further work is needed to  elucidate the dynamics of isospin  under merger conditions.\\
(1) Our most significant assumption is neutrino transparency, which is valid in the limit of a long mean free path for the neutrinos. The behavior of neutrinos in mergers is more complicated, with an energy-dependent mean free path that interpolates between the transparent and trapped regimes, requiring explicit treatment of neutrino transport \cite{Foucart:2022bth}. Our next step will be to use the Urca rate calculation methods developed here to provide state-of-the-art calculations of neutrino-energy-dependent rates, suitable for use in transport computations.\\
(2) Our calculation of the $n\rightleftharpoons p$ rate is based on the standard separation between the direct Urca and modified Urca contributions which are added to give the total rate. Our treatment of direct Urca is full and rigorous, including the entire weak interaction matrix element and integrating the rate over the whole phase space, but the standard expression for the modified Urca rate uses crude approximations, and improving on it is a natural next step \cite{Shternin:2018dcn}.\\
(3) We performed our calculation in the isothermal regime, assuming that the thermal relaxation rate is faster than the dynamical timescales. It would be straightforward to perform a parallel calculation in the adiabatic regime, but previous analyses \cite{Alford:2018lhf,Alford:2022ufz} have found that this does not change the results significantly. This is because the temperatures involved are lower than the Fermi energies of the relevant particles, so the entropy contribution to the pressure is generally a small correction.  In reality thermal conduction in mergers is likely to interpolate between the isothermal and adiabatic regimes, again because of the role of neutrinos, which can have a long mean free path depending on their energy and the ambient density and temperature \cite{Alford:2017rxf}, so they likely dominate the thermal conductivity. \\
(4) Our calculation of bulk viscosity and the damping time for oscillations assumes linear response (``subthermal bulk viscosity'')  where the amplitude of the oscillations is small in the sense that the departure of the chemical potentials from equilibrium is much less than the temperature. Simulations indicate that in the first few milliseconds after merger there are large-amplitude oscillations, for which the suprathermal bulk viscosity \cite{Madsen:1992sx,Reisenegger:2003pd} would be relevant.
We used linear response to obtain suggestive indicators of
the potential importance of isospin relaxation dynamics in mergers. As shown in \cite{Most:2022yhe}, a direct implementation of
the Urca rate equation \eqn{eq:xI-rate-general} in a merger simulation code will naturally incorporate all physical effects, including the subthermal and suprathermal bulk viscosity.
\\
(5) At merger densities we expect nuclear matter
to contain muons, which we neglected in this work. Including them opens up additional equilibration channels ($n\rightleftharpoons p\,\mu$ and purely leptonic processes) leading to a more complicated 
picture with multiple relaxation times \cite{Alford:2022ufz}.
\\
(6) A natural next step is to compare our results with the isospin  equilibration properties of more exotic forms of matter, such as hyperonic \cite{Alford:2020pld} or quark matter \cite{Alford:2006gy,Alford:2008pb,Alford:2010gw,Mannarelli:2009ia,Bierkandt:2011zp}.

In conclusion, this paper provides the most complete treatment to date of the physics of isospin equilibration 
in homogeneous neutrinoless nuclear matter, 
in the density and temperature range that is relevant to neutron star mergers.  Our calculation of the isospin relaxation rate and related phenomena such as bulk viscosity and the damping of density oscillations 
provides a guideline for merger simulators to assess which approximations for isospin equilibration  are appropriate at a given density and temperature, and when an explicit implementation of the relaxation process is required.

\section{Acknowledgements}
We thank Liam Brodie for useful discussions.
MGA and AH are partly supported by the U.S. Department of Energy, Office of Science, Office of Nuclear Physics, under Award No.~\#DE-FG02-05ER41375. ZZ is supported in part by the National Science Foundation (NSF) within the framework of the MUSES collaboration, under grant no.~OAC-2103680.
%\onecolumngrid  
\appendix

\section{Cold $\beta$ equilibrium}
\label{sec:cold-beta-equilibrium}

Previous calculations of the bulk viscosity of neutrinoless matter assumed that isospin  equilibrium occurs when
$\mu_n=\mu_p+\mu_e$, i.e.  $\muIeq=0$ for any $n_B,T$.  
As we have described, this is only valid when the temperature is below about 1\,\MeV. In this low-temperature regime the isospin creation rate depends only on $\De\mu_I$,
\beq
\Ga_I = \dfrac{d\Ga}{d\mu_I} \De\mu_I
 = \dfrac{d\Ga}{d\mu_I} \biggl( 
   \deriv{\mu_I}{x_I}{n_B} \De x_I
 + \deriv{\mu_I}{n_B}{x_I} \De n_B
   \biggr) \, .
\label{eq:Gamma-cold}
\eeq
To make contact with earlier results we define
\beq
 \chinB \equiv -n_B \deriv{\mu_I}{n_B}{x_I}
\label{eq:chinB-def}
\eeq
and
\beq
\chixI \equiv \dfrac{1}{\nbar} \deriv{\mu_I}{x_I}{n_B} \ .
\label{eq:chixI-def}
\eeq
Then comparing \eqn{eq:Gamma-cold} with \eqn{eq:Gamma-expansion} and 
\eqn{eq:xI-rate-linear} we can write $\ga_B$ in terms of $\ga_I$,
\beq
\ba{rl}
\ga_I &= -\chixI \dfrac{d\Ga}{d\mu_I} \, , \\[2ex]
\ga_B &= \dfrac{1}{\bar n_B} \dfrac{\chinB}{\chixI}\,\ga_I \ .
\ea
\label{eq:cold-gammaB}
\eeq
Using the thermodynamic identity \eqn{eq:thermo-identity1}
we can then write the bulk viscosity \eqn{eq:bulk-visc-factorized} as
\beq
\zeta_\text{cold}  = \dfrac{\chinB^2}{\chixI}\,\, \dfrac{\ga_I}{\ga_I^2+\om^2} \ .
\label{eq:zeta-cold}
\eeq
This agrees with previous calculations such as Eq.~(128) of Ref.~\cite{Schmitt:2017efp}.

\section{A relevant thermodynamic identity}
\label{sec:maxwell}

In this appendix we show that
\beq
 \dfrac{1}{n_B} \deriv{p}{x_I}{n_B,T} =  n_B \deriv{\mu_I}{n_B}{x_I,T} \ .
 \label{eq:thermo-identity1}
\eeq
We start by observing that
\beq
 \dfrac{1}{n_B} \deriv{p}{x_I}{n_B,T} = \deriv{p}{n_I}{n_B,T} \ .
 \label{eq:thermo-identity2}
\eeq
Using the thermodynamic expression for pressure,
$p = \mu_B n_B + \mu_I n_I + T s - \ep(n_B,n_I,s)$, and the 
relation
\beq
%\mu_B = \deriv{\ep}{n_B}{n_I} \qquad 
\ba{rl}
\deriv{\ep}{n_I}{n_B,T} &= \deriv{\ep}{n_I}{n_B,s}
+ \deriv{s}{n_I}{n_B,T} \deriv{\ep}{s}{n_B,n_I} \\[2ex]
&= \mu_I + T \,\deriv{s}{n_I}{n_B,T} \, ,
\ea
\eeq
it follows that
\beq
\deriv{p}{n_I}{n_B,T} = n_B \deriv{\mu_B}{n_I}{n_B,T} + n_I \deriv{\mu_I}{n_I}{n_B,T}   \ .
\label{eq:p-derivs}
\eeq
We now use the Maxwell relation
\beq
\deriv{\mu_B}{n_I}{n_B,T}
= \dfrac{\p^2 \ep}{\p n_B \p n_I}\biggr|_T
= \deriv{\mu_I}{n_B}{n_I,T} \ ,
\eeq
where derivatives with respect to $n_B$ are taken at constant $n_I$ and vice versa, so \eqn{eq:p-derivs} becomes
\beq
\ba{rl}
\deriv{p}{n_I}{n_B,T} &= n_B \deriv{\mu_I}{n_B}{n_I,T} +  n_I \deriv{\mu_I}{n_I}{n_B,T} \\[3ex]
&= n_B \deriv{\mu_I}{n_B}{x_I,T} \ ,
\ea
\label{eq:susc-identity}
\eeq
which proves Eq.~\eqn{eq:thermo-identity2} and therefore Eq.~\eqn{eq:thermo-identity1}

\clearpage
\twocolumngrid
\bibliographystyle{JHEP}
\bibliography{reflist}

\end{document}